\documentclass[conference]{IEEEtran}
\IEEEoverridecommandlockouts

\usepackage{cite}
\usepackage{amsmath,amssymb,amsfonts}
\usepackage{multirow,epstopdf}
\usepackage{algorithmic}
\usepackage{graphicx}
\usepackage{textcomp}
\usepackage{xcolor}
\usepackage{booktabs}
\usepackage{threeparttable}
\def\BibTeX{{\rm B\kern-.05em{\sc i\kern-.025em b}\kern-.08em
    T\kern-.1667em\lower.7ex\hbox{E}\kern-.125emX}}

\title{SoCov: Semi-Orthogonal Parametric Pooling of Covariance Matrix for Speaker Recognition}

\author{\IEEEauthorblockN{Rongjin Li\textsuperscript{\dag}, Weibin Zhang\textsuperscript{\dag}, Dongpeng Chen}
\IEEEauthorblockA{\textit{VoiceAI Technologies, Co. Ltd.} \\
Shenzhen, China \\
{\{rongjin,weibin,dongpeng\}@voiceaitech.com}}

\and

\IEEEauthorblockN{Jintao Kang\textsuperscript{$\star$}}
\IEEEauthorblockA{\textit{Institute of Forensic Science,} \\
\textit{Ministry of Public Security} \\
China \\
kangjintao@cifs.gov.cn}

\and

\IEEEauthorblockN{Xiaofen Xing}
\IEEEauthorblockA{\textit{South China University of Technology} \\
Guangzhou, China \\
xfxing@scut.edu.cn}

\thanks{$^\dag$ Both authors contributed equally.}\thanks{$\star$ is the corresponding author.}
}

\begin{document}
\small
\maketitle

\begin{abstract}
In conventional deep speaker embedding frameworks, the pooling layer aggregates all frame-level features over time and computes their mean and standard deviation statistics as inputs to subsequent segment-level layers. Such statistics pooling strategy produces fixed-length representations from variable-length speech segments. However, this method treats different frame-level features equally and discards covariance information. In this paper, we propose the Semi-orthogonal parameter pooling of Covariance matrix (SoCov) method. The SoCov pooling computes the covariance matrix from the self-attentive frame-level features and compresses it into a vector using the semi-orthogonal parametric vectorization, which is then concatenated with the weighted standard deviation vector to form inputs to the segment-level layers. Deep embedding based on SoCov is called ``sc-vector''. The proposed sc-vector is compared to several different baselines on the SRE21 development and evaluation sets. The sc-vector system significantly outperforms the conventional x-vector system, with a relative reduction in EER of 15.5\% on SRE21Eval. When using self-attentive deep feature, SoCov helps to reduce EER on SRE21Eval by about 30.9\% relatively to the conventional ``mean + standard deviation'' statistics.
\end{abstract}


\section{Introduction}

Deep speaker embeddings \cite{b1,b2} have made significant progress, achieving much better recognition accuracy than i-vector \cite{b3}. Based on stacked components such as affine transformations, nonlinear activations, pooling layers and well-designed loss functions, the speaker embedding framework shows a powerful nonlinear modeling ability to discriminate millions of speakers.

In off-the-shelf processing logic, variable-length speech utterances need to be firstly represented as fixed-dimensional vectors. The current de-facto standard is x-vector \cite{b4}. The x-vector method incorporates a statistics pooling layer that receives multiple frame-level deep features (i.e. outputs of the frame-level network), aggregates them along the time axis, and computes the mean and standard deviation. These statistics are concatenated together and passed to subsequent layers. Finally, x-vectors are extracted from an affine component in one of the segment-level hidden layers.

In addition to investigating various network architectures \cite{b5}, loss functions \cite{b6} and adaptation techniques \cite{b7}, different pooling strategies have also been shown to produce much better speaker representations. Among them, an intuitive strategy is to assign different weights to different frame-level deep features instead of treating them equally. This method can highlight vital parts and reduce irrelevant parts. The attention mechanisms \cite{b8} are widely developed, including self-attentive pooling (SAP) \cite{b9}, attentive statistics pooling (ASP) \cite{b10}, self multi-head attention pooling (MHAP) \cite{b11}, etc. However, most attentive pooling functions merely try to improve the system performance based on the traditional mean and standard deviation statistics. The effectiveness of currently used pooling statistics has not been well studied.

The authors in \cite{b12} believe that meta-information, such as speaking style, session, lexical content, etc., captured by temporal pooling strongly depends on the statistics parameters. Except the commonly used first-order mean and second-order standard deviation, they proposed to evaluate two new statistics: skewness and kurtosis (i.e., the 3$^{rd}$ and 4$^{th}$ moments). In the computer vision community, researchers have found that global covariance pooling that is used to replace global average pooling for aggregating the last outputs of deep convolutional neural networks has achieved remarkable performance gains on a variety of vision tasks (\cite{b13}, SMSO in \cite{b14}, iSQRT-COV in \cite{b15}).

In this work, we propose to incorporate the covariance information among different attentive deep features into the pooling layer. The idea of incorporating other second order statistics has been explored before \cite{b16, b17}. Global covariance pooling (GCP) has been explored for speaker recognition in \cite{b16}. \cite{b17} explores channel-wise correlation and thus its effectiveness depends heavily on the underlying network architecture. However, both \cite{b16} and \cite{b17} use methods similar to matrix-flattened for vectorization, which can lead to inputs at the first segment-level layer being excessively large and redundant. Therefore, we will explore how to efficiently utilize the attentive covariance matrix and produce compact and high-fidelity pooling features.

\begin{figure*}
  \centering
  \includegraphics[width=17cm]{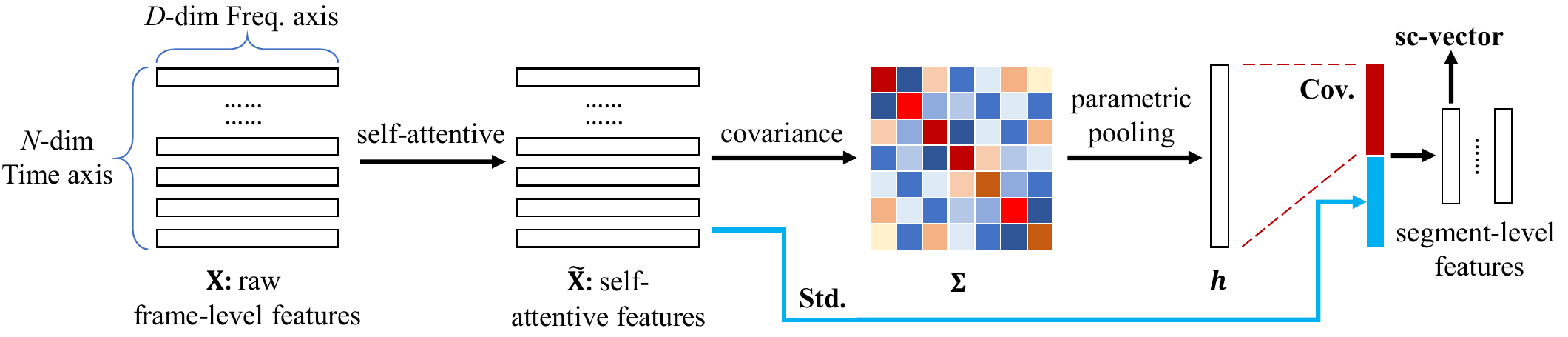}\\
  \caption{Diagram of the semi-orthogonal parametric pooling of self-attentive covariance matrix. The standard deviation and covariance matrix are firstly computed by using the attentive frame-level deep features. The covariance matrix is then vectorized and concatenated with the standard deviation to form inputs for the segment-level network.}\label{fig1}
\end{figure*}

Since covariance matrices are usually very large in typical setups, vectorizing them into high-quality, nearly lossless vectors is a common practice. To alleviate loss of information during compression, previous methods rely heavily on Eigen decomposition (EIG) or singular value decomposition (SVD), which are computationally inefficient on devices and time consuming. We propose to efficiently vectorize the symmetric covariance matrix by multiplying it with a trainable weight vector. The trainable weight vector, which is fixed during inference, is subjected to a semi-orthogonal constraint to preserve more valuable cross time-frequency information during training. By using a trainable parameter layer and semi-orthogonal constraints to compress the attentive covariance matrix, the produced second-order statistics are more representative. A deep embedding based on the SoCov algorithm is called ``sc-vector''.


\section{Related Work}

\subsection{Statistics pooling}

In \cite{b2}, David et al. proposed a statistics pooling (SP) layer to convert frame-level deep features (i.e. outputs of the frame-level network) into segment-level features. Suppose the outputs of frame-level network are $N$ frames of $D$-dimensional deep features, i.e., $\mathbf{X} = \left\{\mathbf{x}_1, \mathbf{x}_2, ..., \mathbf{x}_n, ..., \mathbf{x}_N \right\}$, where $\mathbf{x}_n \in \mathbb{R}^{D}$, thus the size of $\mathbf{X}$ is $N \times D$. The SP layer computes the mean vector $\boldsymbol{\mu}$ and the standard deviation vector $\boldsymbol{\sigma}$ over time as:

\begin{equation}\label{equ1}
    \boldsymbol{\mu} = \frac{1}{N} \sum_{n=1}^N \mathbf{x}_n
\end{equation}

\begin{equation}\label{equ2}
    \boldsymbol{\sigma} = \sqrt{\frac{1}{N} \sum_{n=1}^N (\mathbf{x}_n - \boldsymbol{\mu})^2}
\end{equation}

\begin{equation}\label{equ3}
    \mathcal{P}_{SP} (\mathbf{X}) = concatenate\left\{\boldsymbol{\mu}, \boldsymbol{\sigma}\right\}
\end{equation}

The statistics pooling layer concatenates both $\boldsymbol{\mu}$ and $\boldsymbol{\sigma}$ to produce the inputs of segment-level layers.

\subsection{Self-Attentive pooling}

The authors in \cite{b9} proposed to calculate frame-level weights through a self-attention mechanism. The calculated frame-level weights are incorporated into the statistics pooling layer thus called the self-attentive pooling (SAP) layer. The SAP layer computes a normalized weight matrix $\mathbf{A}$ with the attention parameter matrixes:

\begin{equation} \label{equ4}
	\mathbf{A} = softmax( f (\mathbf{X}\mathbf{W_1})\mathbf{W_2})
\end{equation}

\noindent where $\mathbf{W}_1$ is a trainable matrix of size $D \times D_h$; $\mathbf{W}_2$ is another trainable matrix of size $D_h \times D_r$ and $D_r$ represents the number of attention heads. $f(\cdot)$ is a non-linear activation function and $tanh$ is chosen here. $Softmax(\cdot)$ is performed column-wise. When $D_r > 1$, each attention head of SAP may represent an aspect of discriminative speaker characteristics in the given speech segment.

To simplify the following description, $D_r$ is restricted to be $1$. When $D_r$ is $1$, $\mathbf{A}$ is a weight vector and the corresponding weighted $\boldsymbol{\tilde{\mu}}$ and $\boldsymbol{\tilde{\sigma}}$ can be obtained through:

\begin{equation} \label{equ5}
	\boldsymbol{\tilde{\mu}} = \sum^N_{n=1} \mathbf{x}_n a_n
\end{equation}

\begin{equation} \label{equ6}
	\boldsymbol{\tilde{\sigma}} = \sqrt{\sum^N_{n=1} a_n \mathbf{x}_n \odot \mathbf{x}_n - \boldsymbol{\tilde{\mu}} \odot \boldsymbol{\tilde{\mu}}}
\end{equation}

\noindent where $\odot$ represents the Hadamard product and $a_n$ is the $n^{th}$ elements of the weight vector $\mathbf{A}$.

\section{Proposed Methods}
\label{sec:proposed}

The flowchart of the proposed method, SoCov pooling method, is shown in Fig.1. It firstly transforms the raw frame-level deep features of a speech segment into self-attentive features, i.e. $\mathbf{\widetilde{X}}= \left\{a_1 \mathbf{x}_1, a_2 \mathbf{x}_2, ..., a_n \mathbf{x}_n, ..., a_N \mathbf{x}_N \right\}$. Then the pooling function calculates the covariance matrix $\boldsymbol{\Sigma}$ and compress it with a trainable parametric vector. Details will be described below.

\subsection{Covariance Statistics}
\label{sec:cov}

Given the self-attentive deep features $\mathbf{\widetilde{X}} \in \mathbb{R}^{N \times D}$, the covariance matrix $\boldsymbol{\Sigma} \in \mathbb{R}^{D \times D}$ is computed as:

\begin{equation}\label{equ7}
\boldsymbol{\Sigma} = \sum^N_{n=1}(\mathbf{x}_n a_n - \boldsymbol{\tilde{\mu}})^T (\mathbf{x}_n a_n - \boldsymbol{\tilde{\mu}}) = \mathbf{\overline{X}}^T \mathbf{\overline{X}}
\end{equation}

\noindent where $\mathbf{\overline{X}}$ denotes the mean-subtracted data matrix. The covariance matrix $\boldsymbol{\Sigma}$ is a symmetric positive semi-definite matrix.

\subsection{Parametric Pooling}
\label{sec:semi}

\begin{table*}[]
\centering
\caption{Performance comparison of different pooling methods for FTDNN systems on SRE21 CTS Challenge.} \label{table1}
\begin{threeparttable}
\begin{tabular}{c|c|cc|cc}
\hline
\multirow{2}{*}{Index} & \multirow{2}{*}{FTDNN Systems} & \multicolumn{2}{c|}{SRE21Dev} & \multicolumn{2}{c}{SRE21Eval} \\ \cline{3-6}
                          &             & EER(\%)    & min-Cost   & EER(\%)   & min-Cost     \\ \hline
1	&	$mean$                           & 11.19      & 0.637	& 10.42	   & 0.623	\\
2   &   $cov$-$vec$ (SVD)     	         & 10.28		  & 0.652   & 9.70      & 0.619    \\
3   &   $cov$-$vec$ (SMSO in \cite{b14})    & 10.62		  & 0.774   & 10.65     & 0.738     \\
4   &   $cov$-$vec$ (GCP in \cite{b16})      & 10.44  	  & 0.685   & 9.42      & 0.606        \\
5   &	$cov$-$vec$\tnote{1}				 & 10.15	 	 & 0.711	  & 9.43	       & 0.648	\\
6   &	$cov$-$vec$\tnote{2}				 & 9.27	     & 0.628	  & 8.01	       & 0.577	\\ 
7	&	$standard$ $deviation$				 & 8.04	     & 0.555   & 8.01	   & 0.541	\\ \hline \hline

8   &	$mean$ + $standard$ $deviation$ (SP in \cite{b2})	& 7.90    & 0.539	   & 7.70	   & 0.521	\\ 
9   &   channel-wise correlations in \cite{b17}  & 7.05      & 0.539   & 6.86      & 0.522      \\
10  &   $mean$ + $cov$-$vec$\tnote{1}      & 8.11       & 0.607   & 7.97       & 0.524   \\
11  &	$standard$ $deviation$ + $cov$-$vec$\tnote{1}  & 7.08	& 0.639	& 7.15	   & \textbf{0.515} \\ 
12  &   $mean$ + $cov$-$vec$\tnote{2}     & 7.24        & 0.540   & 7.13       & 0.538  \\
13  &	\textbf{SoCov}: $standard$ $deviation$ + $cov$-$vec$\tnote{2}	& \textbf{6.90}	& \textbf{0.511}	& \textbf{6.51}	& 0.526 \\ \hline \hline

14	&	$mean$ + $standard$ $deviation$	(SAP in \cite{b9})	& 6.09 & 0.494 & 6.34 & 0.504	\\
15	&	$standard$ $deviation$ + $cov$-$vec$\tnote{1} \  (SAP)	& 4.70 & 0.440 & 5.20 & 0.415 \\
16	&	\textbf{SoCov}: $standard$ $deviation$ + $cov$-$vec$\tnote{2} \ (SAP)	& \textbf{3.83} & \textbf{0.361} & \textbf{4.38} & \textbf{0.396} \\ \hline
\end{tabular}
\begin{tablenotes}
        \footnotesize
        \item[1] The covariance matrix is compressed by the parametric pooling without constraint.
        \item[2] The covariance matrix is compressed by the parametric pooling with constraint.
      \end{tablenotes}
    \end{threeparttable}
\end{table*}

In the speaker embedding framework, the covariance matrix $\boldsymbol{\Sigma}$ usually is very large, e.g., $1500 \times 1500$. Therefore compressing the covariance matrix into a compact vector $\mathbf{h} \in \mathbb{R}^{D \times 1}$ is a vital step for subsequent segmentation-level processing of the inputs. Matrix square root normalization has proven to be effective in many computer vision tasks. Existing methods depend heavily on eigen decomposition (EIG) or singular value decomposition (SVD) for computing matrix square root. However, both EIG and SVD are not GPU friendly. In this paper, we employ a different strategy to compress the information of the covariance matrix into a compact vector.

We propose to construct a trainable parameter $\mathbf{w} \in \mathbb{R}^{D \times 1}$ for vectorizing the covariance matrix through $\mathbf{h} = \boldsymbol{\Sigma} * \mathbf{w}$. In the process, we need to ensure that 1) $\mathbf{h}$ contains valuable information extracted from the covariance matrix; 2) $\mathbf{w}$ can be updated to improve the speaker classification accuracy.

To make $\mathbf{h}$ contains valuable information from $\boldsymbol{\Sigma}$, we enforce $\mathbf{h} \mathbf{h}^T = \boldsymbol{\Sigma}\boldsymbol{\Sigma}^T$. As $\mathbf{h} = \boldsymbol{\Sigma} * \mathbf{w}$, we can easily get:

\begin{equation}\label{equ8}
\begin{aligned}
\boldsymbol{\Sigma}\mathbf{w}\mathbf{w}^T\boldsymbol{\Sigma}^T = \boldsymbol{\Sigma}\boldsymbol{\Sigma}^T \\
 \end{aligned}
\end{equation}

\noindent Therefore

\begin{equation}\label{equ9}
\begin{aligned}
\mathbf{w} * \mathbf{w}^T = \mathbf{I}
 \end{aligned}
\end{equation}

As can be seen, $\mathbf{w}\mathbf{w}^T$ will be pushed as close to $\mathbf{I}$ as possible. This is similar to the semi-orthogonal constraint used in \cite{b18}. Thus, we define the following loss function $\mathcal{F}$:

\begin{equation}\label{equ10}
\mathcal{F} \equiv trace(\mathbf{Q}\mathbf{Q}^T) = trace((\mathbf{w}\mathbf{w}^T - \mathbf{I})(\mathbf{w}\mathbf{w}^T - \mathbf{I})^T)
\end{equation}

\noindent where we define $\mathbf{Q} = (\mathbf{w}\mathbf{w}^T - \mathbf{I})$ for convenience. As $\mathbf{Q}$ is symmetric, we can simplify the above formulation without the transposing. To minimize the loss function $\mathcal{F}$, we need to partially differentiate $\mathcal{F}$ with respect to $\mathbf{w}$. By using the chain rule, we can get:

\begin{equation}\label{equ11}
\frac{\partial\mathcal{F}}{\partial \mathbf{w}} = \frac{\partial \mathcal{F}}{\partial \mathbf{Q}}\frac{\partial \mathbf{Q}}{\partial \mathbf{w}} = 4\mathbf{Q}\mathbf{w}
\end{equation}

During the back-propagation stage, the parametric layer $\mathbf{w}$ is updated as follows:

\begin{equation}\label{equ12}
\begin{aligned}
\mathbf{w} \leftarrow \mathbf{w} - \lambda \frac{\partial\mathcal{F}}{\partial \mathbf{w}} = \mathbf{w} - 4\lambda \mathbf{Q}\mathbf{w} \\ = \mathbf{w} - 4\lambda(\mathbf{w}\mathbf{w}^T - \mathbf{I})\mathbf{w}
\end{aligned}
\end{equation}

\noindent where $\lambda$ is the learning rate.

We also hope that $\mathbf{w}$ can be updated to improve the speaker classification accuracy. This can be achieved by using the vector $\mathbf{h}$ with/without other statistics as new inputs for the segment-level network. $\mathbf{w}$ will be trained with the speaker recognition task where cross-entropy is used as the loss function, i.e.,

\begin{equation}\label{equ13}
\mathcal{L}_{CE} = \sum_y l_y log(f(y:\mathbf{\Theta}, \mathbf{w}))
\end{equation}
where $l_y$ denotes the label for the input $y$ and $f(y:\mathbf{\Theta}$, $\mathbf{w})$ represents the output of the whole network with parameters $\mathbf{\Theta}$ and the new added trainable weight $\mathbf{w}$.

Therefore, $\mathbf{w}$ is updated by using both Eq.(12) and Eq.(13). In our implementation, $\lambda$ in Eq.(12) is set to $\frac{1}{8}$ to achieve quadratic convergence \cite{b18}. 

During inference, the covariance vector $\mathbf{h}$ is concatenated with/without other statistics to form the inputs of subsequent segment-level layers. In our experiments and also found in \cite{b16}, it is better to use second-order statistics that produce more comprehensive representation. The proposed sc-vector is based on using both the weighted standard deviation and the covariance vector, i.e.,

\begin{equation}\label{equ14}
\mathcal{P}_{SoCov} (\mathbf{X}) = concatenate\left\{\boldsymbol{\tilde{\sigma}},  \mathbf{h}^T \right\}
\end{equation}

Finally, the dimension of $\mathcal{P}_{SoCov} (\mathbf{X})$ will be reduced at the subsequent segment-level hidden layers, and the sc-vector is extracted from one of the affine layers. The semi-orthogonal parametric pooling method can be easily integrated into a deep neural network with any kinds of architecture and the training process can be achieved by using back-propagation.

\section{Experimental configuration}

All experiments were carried out with the factorization time-delay neural networks (FTDNNs) \cite{b18}, ResNet34 \cite{b19}, and ECAPA-TDNN \cite{b20}. The effectiveness of the proposed methods is evaluated on the SRE 2021 audio track \cite{b21} challenges, including SRE21Dev and SRE21Eval. We computed equal error rate (EER) and minimum detection cost (min-Cost) to evaluate different systems using the official NIST scoring tool.

\subsection{Training Setup}

The NIST SRE CTS Superset \cite{b22}, Vox-Celeb 1 \& 2 \cite{b23}, CN-Celeb 1 \& 2 \cite{b24}, partial Multilingual LibriSpeech (MLS) \cite{b25}, SRE16 and SRE18 were used to train the neural networks. We did data augmentation with additive noise and reverberation by using MUSAN \cite{b26} and RIRS\_NOISES \cite{b27}. Meanwhile, since encoding-decoding of speech is lossy during communication and storage, the MP3 codec was used to simulate such a process. Finally, utterances that are shorter than 5 seconds and speakers with fewer than 8 utterances were all discarded, resulting in a total of 20,563 speakers and 3,589,037 utterances.

All speech was down-sampled to 8KHz if the original recordings were 16KHz speech. The dimension of log Mel filter-banks (F-bank) was set to 64. All the features were extracted every 10ms with a 25ms shift window. The valid frequency was limited to 20-3800Hz. We applied the energy-based voiced activity detection (VAD) and cepstral mean-normalization (CMN) with a sliding window of up to 3 seconds on these acoustic features.

\subsection{Models Setup}

The FTDNN, ResNet34, and ECAPA-TDNN were trained by PyTorch \cite{b28}, while acoustic features and back-end models were implemented with Kaldi \cite{b29}. As the main experimental backbone, the FTDNN (without any recurrent layers) is described in \cite{b30}. The embeddings (x-vector/sc-vector) are extracted after the first affine layer of the segment-level network. As for the loss function, we used the boundary discriminative AM-Softmax to minimize intra-speaker variation and maximize inter-speaker discrepancy \cite{b6}. The margin was set to 0.35 and the scale was 64. The above settings are the same in the ResNet34 and ECAPA-TDNN.

All network models were trained using the SGD optimizer and the exponential learning rate strategy. The weight decay of SGD was $3e^{-4}$. The maximum and minimum learning rates were set to $1e^{-2}$ and $1e^{-5}$, respectively. We trained the models for 4 epochs with a batch-size of 128 and a chunk-size of 4 seconds. Then the 512-dimensional embeddings were extracted for scoring. The PCA and LDA reduced the dimension of embeddings from 512 to 150, and from 150 to 100, respectively. We trained a PLDA model \cite{b31} for scoring speaker recognition. In order to ensure the reproducibility of the ablation analyses, we prohibited any fine-tuning, domain adaptation, or score normalization to correctly assess the contribution of the proposed SoCov.

\section{Experimental results}
\label{sec:res}

The performance of FTDNN systems is listed in Table 1. We first assessed each single statistic, and their results are shown in the first seven rows. As for the covariance statistic, we evaluated covariance matrix vectorization (called $cov$-$vec$) with various compression methods, including SVD, SMSO, GCP, without and with the semi-orthogonal constraint in Eq.(9). The results of this series show that the amount of information carried by the simple mean statistic is limited, and high-order statistics are more important for speaker recognition. The SVD is more computationally heavy while training and its gradients are unstable. Both SMSO and GCP simulate the SVD process, where their training is fast but compression is lossy. Under supervised learning and semi-orthogonal constraints, our proposed $cov$-$vec$ outperforms SVD pooling and approaches the effect of standard deviation.

\begin{figure}
  \centering
  \includegraphics[width=7.5cm]{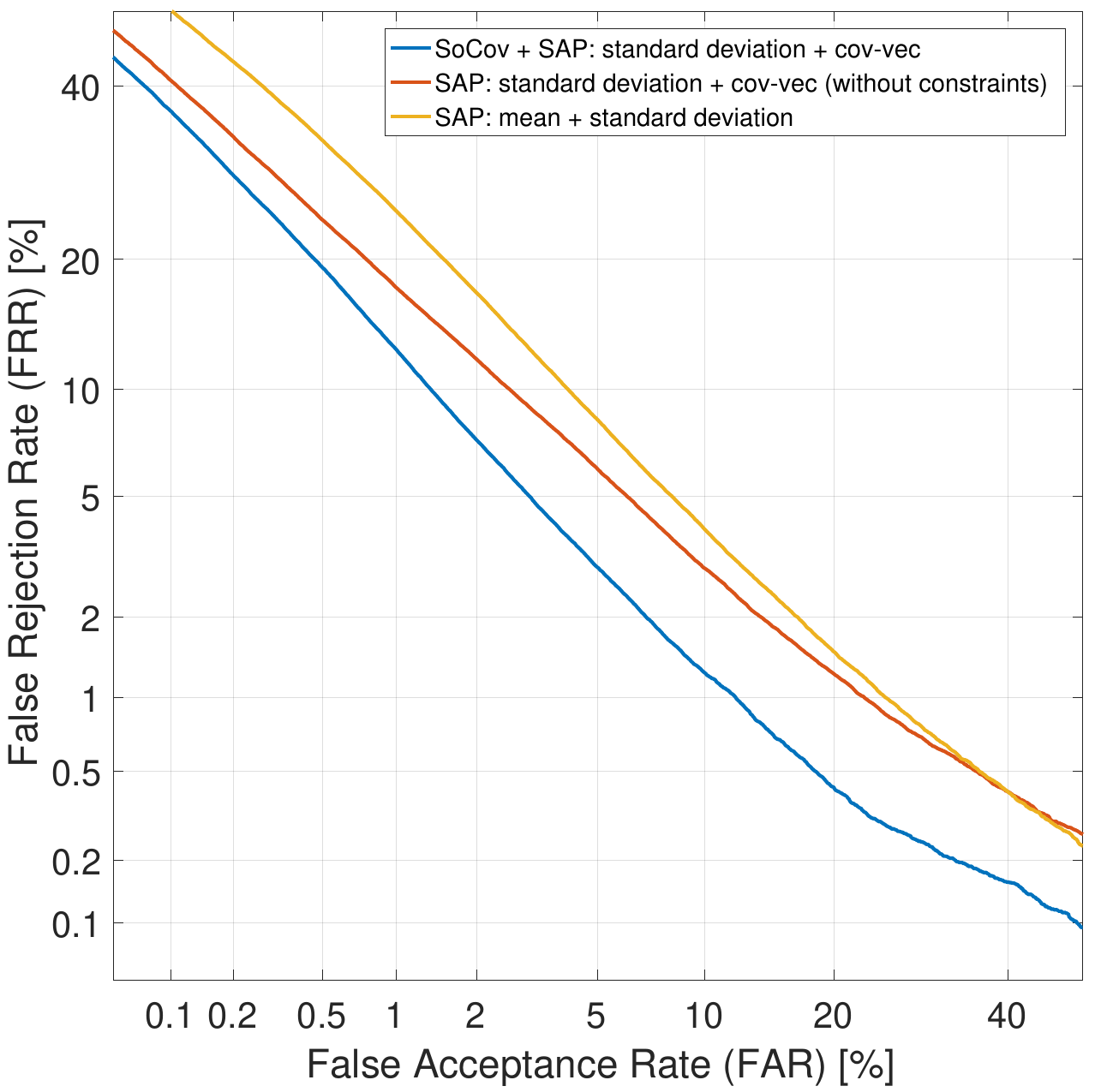}\\
  \caption{The DET curves of self-attentive pooling (SAP) FTDNN systems with different statistics on NIST SRE21Eval.}\label{fig2}
\end{figure}

We then went on to evaluate the effectiveness of combining different statistics. Two baseline systems, the SP layer and channel-wise correlations layer (CWC) are at the 8-$th$ and 9-$th$ row, respectively. The CWC is the same as the best configuration in \cite{b17}, and its motivation is similar to the one we proposed. Clearly, with constraints our proposed $cov$-$vec$ are more effective in the different combinations. Specially, combining two high-order statistics, i.e., SoCov, is more meaningful and powerful in discriminating speakers. Compared with the SP, the SoCov system relatively reduces EER by 12.7\% on SRE21Dev and 15.5\% on SRE21Eval. Given the dimensionality extension of combining the three statistics, we use only two statistics for splicing to align the dimensional variables.

Subsequently, we evaluated systems with self-attentive pooling (SAP). We set $D_h = floor(D / 4)$ and $D_r = 1$ for the SAP hyper-parameters. The proposed SoCov achieves the best performance on both SRE21Dev and SRE21Eval. On SRE21Eval, the proposed SoCov + SAP system achieves 4.38\% EER and 0.396 minimum cost. Compared with the baseline SP system, our proposed system relatively reduces EER by about 30.9\% on SRE21Eval.

To give a more intuitive comparison between the sc-vector and x-vector systems based on self-attentive deep features, we draw the DET curves of self-attentive pooling (SAP) systems with different statistics on NIST SRE21Eval in Fig.2. Obviously, the proposed sc-vector performs the best at all operating points.

\subsection{Effectiveness of constraints}

From the results of covariance matrix vectorization with/without semi-orthogonal constraints, we observe that the proposed constraint for covariance matrix vectorization constantly improve the performance. When self-attentive pooling (SAP) is used, the proposed constraint help reduce EER by 18.5\% on SRE21Dev and 15.8\% on SRE21Eval respectively and relatively. Obviously, the proposed algorithm effectively helps preserve salient information inside the covariance matrix to discriminate different speakers.

\begin{table}[]
\caption{Performance comparison of different pooling methods for ResNet34 and ECAPA-TDNN systems on SRE21 CTS Challenge.} \label{table2}
\begin{tabular}{ccccc}
\hline
\multirow{2}{*}{\begin{tabular}[c]{@{}c@{}}ResNet34\\ Systems\end{tabular}}   & \multicolumn{2}{c}{SRE21Dev} & \multicolumn{2}{c}{SRE21Eval} \\ \cline{2-5} 
               & EER(\%)      & min-Cost      & EER(\%)       & min-Cost      \\ \hline
SAP            & 7.54         & 0.655         & 6.31          & 0.551          \\
SoCov          & 7.19         & 0.529         & 6.22          & \textbf{0.471}    \\
SoCov + SAP    & \textbf{4.79}  & \textbf{0.402}  & \textbf{5.05}   & 0.532           \\ \hline \hline

\multirow{2}{*}{\begin{tabular}[c]{@{}c@{}}ECAPA-TDNN\\ Systems\end{tabular}} & \multicolumn{2}{c}{SRE21Dev} & \multicolumn{2}{c}{SRE21Eval} \\ \cline{2-5} 
              & EER(\%)      & min-Cost      & EER(\%)       & min-Cost      \\ \hline
SAP          & 6.40          & 0.524         & 5.88           & 0.501            \\
SoCov        & 6.78          & 0.481         & 6.13           & 0.449           \\
SoCov + SAP   & \textbf{4.03}  & \textbf{0.383}  & \textbf{4.71}  & \textbf{0.426}    \\ \hline
\end{tabular}
\end{table}

\subsection{Effectiveness on Other Backbones}

Finally, in order to fully test the SoCov method, we verified it in ResNet34 and ECAPA-TDNN backbones. The ECAPA-TDNN systems perform much better than the ResNet34, suggesting that the performance of ECAPA-TDNN under cross-domain testing conditions is promising. 

When using SoCov + SAP in the ECAPA-TDNN, it is more outstanding results on SRE21. On SRE21Eval, the SoCov + SAP system achieves 4.71\% EER and 0.426 min-Cost in Table 2. Above result demonstrates our proposed SoCov approach is able to improve the performance of different backbones.

\section{Conclusion}
\label{sec:con}

The conventional x-vector uses the mean and standard deviation as the statistics in the pooling layer, where the correlation among different random variables is discarded. We propose the sc-vector architecture that incorporates both the standard deviation and information from the covariance matrix. In addition, we propose an algorithm to effectively compress the covariance matrix into a vector in order to substantially reduce the input dimension of the following segment-level network. Experiments show that the sc-vector (``\textbf{s}tandard deviation + \textbf{c}ov-vec'') significantly outperforms the conventional x-vector (``mean + standard deviation''), especially when self-attentive deep features are used.

\section*{Acknowledgment}

This work was supported in part by 2019JB032, Nansha key project 2022ZD011 and 2021YFF0600202-YF-ZLJC2102-2-1.

\vspace{12pt}

\end{document}